\documentclass[11pt,a4paper]{article}
\usepackage{dsfont}
\usepackage{slashed}
\usepackage{setspace}
\pdfoutput=1
\usepackage[utf8]{inputenc}
\usepackage{a4wide}
\usepackage{cite}
\usepackage{xcolor}
\usepackage{amssymb}
\usepackage{amsfonts}
\usepackage{graphicx}
\usepackage{mathbbol}
\usepackage{amssymb}
\usepackage{booktabs}
\usepackage{colortbl}
\usepackage{collcell}
\usepackage{tabularx}
\usepackage{multirow}
\usepackage{appendix}
\DeclareSymbolFontAlphabet{\amsmathbb}{AMSb}
\usepackage{amsfonts}
\usepackage{amssymb}
\usepackage[centertags]{amsmath}

\usepackage[normalem]{ulem}
\newcommand{\virg}[1]{``#1''}

\definecolor{myred}{cmyk}{0,1,1,0.55}
\definecolor{mygreen}{rgb}{0.27, 0.64, 0.48}
\definecolor{mygray}{gray}{.95}
\usepackage[colorlinks=true,urlcolor=myred,linkcolor=black,citecolor=mygreen]{hyperref}

\def\sph{\text{sph}}

\def\ltap{\ \raisebox{-.4ex}{\rlap{$\sim$}} \raisebox{.4ex}{$<$}\ }
\def\gtap{\ \raisebox{-.4ex}{\rlap{$\sim$}} \raisebox{.4ex}{$>$}\ }

\allowdisplaybreaks[1]
\usepackage{a4wide}
\usepackage{colordvi}

\newcommand{\bec}{\begin{cases}}
\newcommand{\eec}{\end{cases}}
\newcommand{\beq}{\begin{equation*}}
\newcommand{\eeq}{\end{equation*}}
\newcommand{\be}{\begin{equation}}
\newcommand{\ee}{\end{equation}}
\newcommand{\ba}{\begin{eqnarray}}
\newcommand{\ea}{\end{eqnarray}}

\begin{document}
\begin{titlepage}

${}$\vskip 3cm
\vspace*{-15mm}
\begin{flushright}
 SISSA 10/2022/FISI\\
\end{flushright}
\vspace*{0.7cm}

\vskip 9mm
\begin{center}
{\bf\Large Tests of Low-Scale Leptogenesis in
Charged Lepton Flavour\\
\vspace{2mm}
Violation Experiments
} \\
[5mm]
\renewcommand*{\thefootnote}{\fnsymbol{footnote}}
A.~Granelli$^{~a,b,c}$ \footnote{\href{mailto:agranell@sissa.it}{agranell@sissa.it}},
J.~Klari\'c$^{~d}$ \footnote{\href{mailto:juraj.klaric@uclouvain.be}{juraj.klaric@uclouvain.be}}
and S.~T.~Petcov$^{~a,b,e}$
\renewcommand*{\thefootnote}{\arabic{footnote}}
\setcounter{footnote}{0}\footnote{Also at:
Institute of Nuclear Research and Nuclear Energy,
Bulgarian Academy of Sciences, 1784 Sofia, Bulgaria.} \\
\vspace{2mm}
$^{a}$\,{\it SISSA, via Bonomea 265, 34136 Trieste, Italy.} \\
$^{b}$\,{\it INFN, Sezione di Trieste, via Valerio 2, 34127 Trieste, Italy.} \\
$^{c}$\,{\it IFPU, via Beirut 2, 34151 Trieste, Italy.}\\
$^{d}$\,{\it Centre for Cosmology, Particle Physics and Phenomenology, Université Catholique de Louvain, Louvain-la-Neuve B-1348, Belgium.}\\
$^{e}$\,{\it Kavli IPMU (WPI), UTIAS, The University of Tokyo, Kashiwa,
Chiba 277-8583, Japan.}\\
\end{center}

\begin{abstract}
We consider low-energy tests of low-scale leptogenesis based on the
type I seesaw scenario with three right-handed
singlet neutrinos $\nu_{l R}$. In this scenario, successful leptogenesis
is possible for quasi-degenerate in mass heavy Majorana neutrinos $N_{1,2,3}$,
$M_{1,2,3}\cong M$, $|M_j - M_i|\ll M$, $i\neq j = 1,2,3$,
heavy Majorana neutrino masses $M \sim (0.05 - 5\times 10^5)$ GeV,
and $N_j$ charged current and neutral current
weak interaction couplings as large as $\mathcal{O}(10^{-2})$.
We derive the constraints on the corresponding
leptogenesis parameter space from the existing data
from low-energy experiments, including the limits from
the experiments on $\mu \rightarrow e \gamma$ decay and
on the rate of $\mu - e$ conversion in gold.
We show also that the planned and upcoming experiments on
charged lepton flavour violation with $\mu^\pm$,
MEG II on the $\mu \rightarrow e\gamma$ decay,
Mu3e on $\mu \rightarrow eee$ decay,
Mu2e and COMET on $\mu - e$ conversion in aluminium
and PRISM/PRIME on $\mu - e$ conversion in titanium,
can probe significant region of the
viable leptogenesis parameter space, and thus
have a potential for a discovery.
Experiments on $\tau \to eee(\mu\mu\mu)$ and $\tau \to e(\mu)\gamma$ decays 
(e.g., BELLE II) also can probe a part of the leptogenesis parameter space,
although a relatively small one.

\end{abstract}

\end{titlepage}
\renewcommand*{\thefootnote}{\arabic{footnote}}
\setcounter{footnote}{0}
\setcounter{page}{2}

\setcounter{section}{0}
%%%%%%%%%%%%%%%%%%%%%%%%%%%%
\section{Introduction}
\label{sec:intr}
%%%%%%%%%%%%%%%%%%%%%%%%%%%%

\indent
In the present article, we investigate the possibility to test the low-scale leptogenesis scenarios of generation of the Baryon Asymmetry of the Universe (BAU)
\cite{Fukugita:1986hr,Kuzmin:1985mm,Pilaftsis:1997jf,Pilaftsis:2003gt,Akhmedov:1998qx, Asaka:2005pn}
based on the type I seesaw mechanism
\cite{Minkowski:1977sc,Yanagida:1979as,GellMann:1980vs,Glashow:1979nm,Mohapatra:1979ia} in experiments sensitive to beyond the Standard Model physics
at sub-TeV scales.  As is well known, an integral part of  the
type I seesaw mechanism and the related leptogenesis
scenarios are the right-handed (RH) neutrinos
$\nu_{aR}$ (RH neutrino fields $\nu_{aR}(x)$),
which can be added as $\text{SU}(2)_{\rm L}$ singlets
to the Standard Model (SM) without modifying its
basic properties. Such a SM extension with two seesaw
RH neutrinos and, correspondingly, with two heavy Majorana
neutrinos $N_{j}$ with definite masses $M_{j} > 0$, $j=1,2$,
is the minimal set-up in which leptogenesis can
be realised, satisfying the three Sakharov's
conditions \cite{Sakharov:1967dj}
for a dynamical generation of the matter-antimatter asymmetry.

In classical thermal leptogenesis with $N_j$
having hierarchical mass spectrum, the generation of the BAU,
due to the out-of-equilibrium L-, C- and CP-violating decays of $N_j$,
takes place at scales which are typically by a few to several orders
of magnitude smaller than the scale of unification of
the electroweak and strong interactions, $M_\text{GUT}\cong 2\times 10^{16}$ GeV
(see, e.g.,~\cite{Granelli:2021fyc}
and the recent review article \cite{Bodeker:2020ghk},
which include also extended lists of references).
The scale of leptogenesis is determined, in general,
by the values and the spectrum of masses of the heavy Majorana
neutrinos $N_j$. A rather detailed analysis of the high
scale thermal (non-resonant)
leptogenesis scenario with three RH neutrinos
performed in \cite{Moffat:2018wke,Moffat:2018smo}
showed that, with flavour effects taken into account
and mildly hierarchical heavy Majorana neutrino masses,
$M_2 \sim 3M_1$, $M_3\sim 3M_2$,
the leptogenesis scale can be as low as $M_1 \sim 10^6$ GeV.
Testing experimentally even this high scale
leptogenesis scenario seems impossible at present.

A unique possibility to test experimentally the leptogenesis idea
is provided by the low-scale scenarios based on the type I seesaw mechanism
proposed in \cite{Pilaftsis:1997jf,Pilaftsis:1998pd,Pilaftsis:2003gt}
and in \cite{Akhmedov:1998qx,Asaka:2005pn}.
In these scenarios, the heavy Majorana neutrinos can have masses
at the sub-TeV scales, which makes the scenarios testable,
in principle, at colliders (LHC and/or future planned)
and/or at low-energy experiments (see further).

In resonant leptogenesis \cite{Liu:1993tg,Flanz:1994yx,Flanz:1996fb,Covi:1996fm,Covi:1996wh,Pilaftsis:1997jf,Buchmuller:1997yu,Pilaftsis:1998pd,Pilaftsis:2003gt},
the baryon asymmetry
is produced exclusively by the CP-violating
$N_{j}$ and Higgs decays mediated by the neutrino Yukawa couplings
with $N_j$ having masses $M_j < (\ll) \,1$ TeV.
In the simplest case with two RH neutrinos,
the resonant regime is realised if
the associated two heavy Majorana neutrinos $N_{1,2}$
form a pseudo-Dirac pair\footnote{It was shown in~\cite{Pilaftsis:1991ug, Kersten:2007vk} that, in this case, the radiative corrections to the light neutrino masses are negligible.
We verify that this condition is satisfied, and reject all points for which the radiative corrections are comparable to the tree level contribution.} \cite{Wolfenstein:1981kw,Petcov:1982ya}
such that the splitting between their masses,
$M_2 - M_1 \equiv \Delta M > 0$,
is of the order of the $N_{1,2}$ decay widths  $\Gamma_{1,2}$:
$\Delta M/\Gamma_{1,2}\sim 1$, which typically implies also that
$\Delta M \ll M_{1,2}$.
This scenario was re-visited using the formalism of Boltzmann equations
most recently in \cite{Granelli:2020ysj},
where the authors concentrated on the case of $M_{1,2} \ltap 100$ GeV,
$\Delta M \ll M_{1,2}$ (for earlier discussions see, e.g.,
\cite{1606.00017, Hambye:2017elz}).
Both the relevant $1\leftrightarrow 2$ decays and inverse decays
and $2\leftrightarrow 2$ scattering processes
(involving quarks and gauge fields),
including flavour effects and thermal effects
(thermal masses and soft collinear processes involving gauge fields
in the thermal plasma), were taken into account.
Results were presented in \cite{Granelli:2020ysj}
for the two possible  $N_{1,2}$ initial abundances at temperature
$T_0 \gg T_\sph$, $T_\sph$ being the sphaleron decoupling temperature
$T_\sph = 131.7$ GeV
\footnote{The baryon asymmetry $\eta_B$ during
 the generation process \virg{freezes} at $T_\sph$ as the temperature
of the Universe decreases and the value of
$\eta_B$ at $T_\sph$ should be compared with the
observed one.}:
i) $N_{1,2}$ Thermal Initial Abundance (TIA), and
ii) $N_{1,2}$  Vanishing (zero) Initial Abundance (VIA).
The light neutrino mass spectrum
with normal ordering (NO)
(see, e.g., \cite{Tanabashi:2018oca})
was considered. It was found that successful
resonant leptogenesis is possible
in the VIA (TIA) case for masses of the heavy Majorana neutrinos
across the whole of the experimentally
accessible region of $M_{1,2} \cong 0.3~(5.0) - 100$ GeV,
and for values of the charged and neutral current
couplings of $N_{1,2}$ in the weak interaction Lagrangian,
denoted in \cite{Granelli:2020ysj}
as $(RV)_{\ell j}$, $\ell=e,\mu,\tau$, $j=1,2$, in the range of
$(10^{-6} - 5\times 10^{-5})$.

In \cite{Akhmedov:1998qx,Asaka:2005pn}, the so-called
\virg{freeze-in} leptogenesis mechanism by which the BAU is generated via RH neutrino oscillations during the epoch when the RH neutrinos, or equivalently, the heavy Majorana neutrinos $N_j$, are being produced and are out of equilibrium, was put forward. This mechanism was extensively studied
(see, e.g., \cite{hepph0605047,1112.5565,canetti2013dark,Shuve:2014zua,1508.03676,Drewes:2016gmt,Hernandez:2016kel,1704.02692,Ghiglieri:2017gjz,Drewes:2017zyw,Abada:2018oly} and references quoted therein).

Resonant leptogenesis and leptogenesis via neutrino oscillations were usually
treated as separate in baryogenesis mechanisms.
Only recently, the parameter space of the two scenarios was studied in
a unified framework in \cite{Klaric:2020lov}
(see also \cite{Klaric:2021cpi}) based on density matrix-like equations
(see, e.g., \cite{Dev:2017wwc,Garbrecht:2018mrp} for a review of
the formal treatments of resonant leptogenesis).
\footnote{The density matrix equations used in this work were derived
independently in the density matrix formalism for mixing
neutrinos~\cite{Sigl:1993ctk,Ghiglieri:2017gjz,Eijima:2017anv} and in
the Closed-Time-Path (CTP) formalism using the gradient
expansion~\cite{Garbrecht:2011aw,Drewes:2016gmt,Antusch:2017pkq}.
The resulting asymmetries were shown to agree with the full CTP approach
in a static Universe, for $\Delta M/M \ll 1$~\cite{Dev:2017wwc}.
On the other hand, following similar considerations, the authors of~\cite{BhupalDev:2014pfm,BhupalDev:2014oar} claim to have found an additional source of CP violation related to the phenomenon of resonant flavour mixing, distinct from that of heavy Majorana neutrino oscillation, which can lead to additional contribution to the baryon asymmetry, and thus further theoretical uncertainty.
}
Considering the case
of two heavy Majorana neutrinos $N_{1,2}$ forming a pseudo-Dirac pair,
in \cite{Klaric:2020lov} it was shown that
i) the observed baryon asymmetry can be generated for
all experimentally allowed values of the Majorana neutrino masses
$M_{1,2} \cong M \gtrsim 100$ MeV and up to
the TeV scale,  and that
ii) leptogenesis is effective in a broad range of
the relevant parameters, including mass splitting between
the two Majorana neutrinos
as large as $\Delta M/M\sim 0.1$, as well as
couplings of $N_{1,2}$ in the weak charged lepton current
which depend on the value of
$M$: for, e.g., $M = 1$ and 50 GeV, they are in the range
of $(10^{-5} - 10^{-3})$ and $(10^{-6} - 3\times 10^{-5})$,
respectively. The results derived in \cite{Klaric:2020lov}
and in \cite{Granelli:2020ysj}
are largely compatible in the leptogenesis parameter space regions
where they can be compared, such as, e.g.,
in the regions corresponding to the
case of TIA and light neutrino mass spectrum with NO.
The region of viable leptogenesis parameter space for $M \gtrsim 0.2$ GeV
found in \cite{Klaric:2020lov,Granelli:2020ysj},
leads to an upper bound on the weak lepton charged current (CC) interactions
$M \cdot U^2 \lesssim 5\cdot 10^{-6}$ GeV, where $U^2 \equiv \sum_{\ell i} |(R V)_{\ell i}|^2$.
This is too small to be probed in low-energy experiments
\footnote{The only exception could be the neutrinoless double beta decay
experiments, which can have a contribution from the heavy Majorana
neutrinos with large mass splittings $\Delta M/M \gtrsim 10^{-3}$,
and masses below $2$ GeV, as was shown
in~\cite{Drewes:2016lqo,Hernandez:2016kel}.},
but could be probed in fixed target experiments~\cite{Agrawal:2021dbo,Abdullahi:2022jlv}, future colliders~\cite{Antusch:2016ejd,FCC:2018evy,CEPCStudyGroup:2018ghi,Abdullahi:2022jlv}, or potentially already at the HL-LHC~\cite{Boiarska:2019jcw,Drewes:2019fou} (see, e.g., Fig.~1 in \cite{Klaric:2020lov}).

The unified treatment of low-scale leptogenesis
was extended in \cite{Drewes:2021nqr} to the case of
three quasi-degenerate heavy Majorana neutrinos $N_{1,2,3}$, with
$M_{1,2,3}\cong M$. The authors
of \cite{Drewes:2021nqr} presented  results for
$M$ between $50$ MeV and $70$ TeV, focusing on the case of light neutrino
mass spectrum with NO, either hierarchical (NH) or quasi-degenerate (QD),
and considered both vanishing and thermal initial conditions.
The major finding in \cite{Abada:2018oly,Drewes:2021nqr} is that the range
of heavy Majorana neutrino CC and neutral current (NC) couplings for which
one can have successful leptogenesis
is by several orders of magnitude larger
than the range in the scenario with two heavy Majorana neutrinos,
reaching at, e.g., $M = 100$ GeV values
$\sim 5\times 10^{-2}$ in the case of TIA and even somewhat larger
values in the  case of VIA.
\footnote{
Note that the possibility of large couplings was
found in~\cite{Pilaftsis:2004xx,Pilaftsis:2005rv} in the special regime
of resonant $\tau$-leptogenesis in which the coupling of the heavy
Majorana neutrinos to the $\tau$
charged lepton is negligible,
while the couplings to $e$ and $\mu$, although relatively large,
do not play a role in leptogenesis.
Although the results of~\cite{Abada:2018oly,Drewes:2021nqr} allow
for such a BAU production
mechanism, this was not found to be the dominant mechanism which is 
% responsible for
associated with large Majorana neutrino couplings
to $e$, $\mu$ and $\tau$ charged leptons.
}
For heavy Majorana neutrinos with masses below the TeV scale,
a large range of couplings can already be probed in direct searches at the
LHC~\cite{Atre:2009rg,Deppisch:2015qwa,Cai:2017mow,Agrawal:2021dbo,Abdullahi:2022jlv},
as well as in fixed target experiments~\cite{Agrawal:2021dbo,Abdullahi:2022jlv} and future colliders~\cite{Antusch:2016ejd,FCC:2018evy,CEPCStudyGroup:2018ghi,Abdullahi:2022jlv}.
In the present article, we investigate the possibility
to test directly the low-scale leptogenesis scenarios
discussed in~\cite{Drewes:2021nqr} (see also \cite{Abada:2018oly})
in upcoming high precision experiments on charged lepton flavour violation (cLFV) searching for  $\mu^\pm \to e^\pm + \gamma$ and
$\mu^\pm \to e^\pm + e^+ + e^-$ decays and for $\mu - e$ conversion in nuclei.

 %%%%%%%%%%%%%%%%%%%%%%%%%%%%
%
\section{Aspects of the Seesaw Formalism and the Analysis}
\label{sec:param}
%
%%%%%%%%%%%%%%%%%%%%%%%%%%%%

In the set-up with
three singlet
RH neutrinos $\nu_{aR}$
and in the leptogenesis framework based on type I seesaw mechanism,
in general, the required non-conservation of the
total lepton charge $L$ is provided, as is well known, by the Majorana mass term of the singlet neutrinos $\nu_{aR}$ and the neutrino
Yukawa coupling ${\cal L}_{\rm Y}(x)$  involving  $\nu_{aR}$
and the SM lepton and Higgs doublets, $\psi_{\ell L}(x)$ and $\Phi(x)$.
The requisite breaking of C- and CP-symmetries
is ensured by the $\nu_{aR}$ Majorana mass term
and/or the Yukawa coupling  ${\cal L}_{\rm Y}(x)$.

In the diagonal mass basis of the RH neutrinos
$\nu_{aR}$ and the charged leptons $\ell^\pm$, $\ell=e,\mu,\tau$,
which proves convenient for the leptogenesis
analysis and was used in
\cite{Klaric:2020lov,Granelli:2020ysj,Drewes:2021nqr},
the neutrino Yukawa coupling ${\cal L}_{\rm Y}(x)$ and
the seesaw Majorana mass term are given by:
%%%%%%%%%%%%%%%%%%%%%%%%%%%%%%%%%%%%
\begin{equation}
\label{eq:Ynu}
{\cal L}_{\rm Y,M}(x) =
-\,\left (Y_{\ell i} \overline{\psi_{\ell L}}(x)\,i\tau_2\,\Phi^*(x)\,N_{iR}(x)
+ \hbox{h.c.} \right )
-\,\frac{1}{2}\,M_{i}\,\overline{N_i}(x)\, N_i(x)\,,
\end{equation}
%%%%%%%%%%%%%%%%%%%%%%%%%%%%%%%%%%%%
%
\noindent
where $Y_{\ell i}$ is the matrix of neutrino Yukawa couplings
(in the chosen basis),
$(\psi_{\ell L}(x))^T = (\nu^T_{\ell L}(x)~~\ell ^T_{L}(x))$, $\ell =e,\mu,\tau$,
$\nu_{\ell L}(x)$ and $\ell_{L}(x)$ being the left-handed (LH) flavour neutrino
and charged lepton fields,
$(\Phi(x))^T = (\Phi^{(+)}(x)~\Phi^{(0)}(x))$
and $N_i$ ($N_i(x)$) is the heavy Majorana neutrino
(field) possessing a mass $M_i > 0$.  In the same basis,
the flavour neutrino fields
$\nu_{\ell L}(x)$, $\ell =e,\mu,\tau$, which enter into the expressions
of the charged and neutral currents
in the weak interaction Lagrangian, are given by:
%%%%%%%%%%%%%%%%%%%%%%%%%%%%%%%%
\begin{equation}
\nu_{\ell L}(x) = \sum_i (1 + \eta)U_{\ell i}\nu_{iL}(x) +
\sum_j (RV)_{\ell j} N_{jL}(x)\,,
\label{eq:nulnuiNj}
\end{equation}
%%%%%%%%%%%%%%%%%%%%%%%%%%%%
%
where $N_{jL}(x)$ are the LH components
of the fields of the heavy neutrinos $N_j$,
$\nu_{iL}(x)$, $i=1,2,3$, are the LH  components
of the fields of three light Majorana neutrinos $\nu_i$
having  masses $m_i$, $m_i \ltap 0.5~{\rm eV} \ll M_j$,
$U$ is a $3\times3$ unitary matrix and
$\eta =  -(1/2)(RV)(RV)^\dagger$.
The matrix $R$ is determined by
$R \cong M_D\, M^{-1}_{N}$, $M_D$ and $M_N$ being the
seesaw neutrino Dirac and the RH neutrino Majorana mass matrices,
respectively, $|M_D| \ll |M_N|$, and
$V$ is the unitary matrix which
(to leading approximation in $M_D/M_N$)
diagonalises the Majorana mass matrix of the heavy RH neutrinos $M_N$
(see, e.g., \cite{Ibarra:2010xw}).
The matrix $M_D$ is related to the matrix of neutrino Yukawa
couplings $Y$ in Eq.~(\ref{eq:Ynu}) as follows:
$M_{D} = (v/\sqrt{2})Y\,V^T$, $v = 246$ GeV. The Majorana mass
matrix of the LH flavour neutrinos is given by the
well known seesaw expression:
%%%%%%%%%%%%%%%%%%%%%%%%%%%%%%%%%%%%%
\be
(m_\nu)_{\ell \ell'} \cong -\, \left[M_{D}\,M^{-1}_{N}\,(M_D)^{T}\right]_{\ell \ell'} =
-\, \dfrac{v^2}{2}\, Y_{\ell j}\,M^{-1}_j\, Y^T_{j \ell'}
= (U\, \hat{m}_\nu\, U^T)_{\ell \ell'}\,,
\label{eq:seesanuMajM}
\ee
%%%%%%%%%%%%%%%%%%%%%%%%%%%%%%%%%%
%
where $\hat{m}_\nu = {\rm diag}(m_1,m_2,m_3)$.

It follows from Eq.~(\ref{eq:nulnuiNj}) that, in the seesaw scenario we are considering, the PMNS matrix has the form:
%%%%%%%%%%%%%%%%%%%%%%%%%%%%%%%
\begin{equation}
U_{\text{PMNS}} = (1+\eta)\,U\,.
\end{equation}
%%%%%%%%%%%%%%%%%%%%%%%%%%%%%%%%
%
The matrix $\eta$ describes
the deviations from unitarity of the PMNS matrix.
The elements of $\eta$
are constrained by electroweak data and data
on flavour observables \cite{Fernandez-Martinez:2015hxa,Blennow:2016jkn}.
For $M_j \gtap 500$ MeV and depending on the element of $\eta$,
these constraints are in the range $(10^{-4} - 10^{-3})$
at $2\sigma$ C.L. For $M_j$ larger than the electroweak scale,
the constraint on $\eta_{e\mu} = \eta_{\mu e}$ is even stronger:
$|\eta_{e\mu}| < 1.2\times 10^{-5}$.
Given the stringent
upper bounds on the elements of $\eta$,
to a very good approximation one has:
$U_{\text{PMNS}} \cong U$. Following
\cite{Klaric:2020lov,Granelli:2020ysj,Drewes:2021nqr}
we  use in our analysis the standard parametrisation
of the PMNS matrix $U_\text{PMNS}$ \cite{Tanabashi:2018oca}:
%%%%%%%%%%%%%%%%%%%%%%%%%%%%
\begin{equation}
\label{PMNS}
U_\text{PMNS} = \begin{pmatrix}
c_{12}c_{13}&s_{12}c_{13}&s_{13}\text{e}^{-i\delta}\\
-s_{12}c_{23}-c_{12}s_{23}s_{13}\text{e}^{i\delta}&c_{12}c_{23}-s_{12}s_{23}s_{13}\text{e}^{i\delta}&s_{23}c_{13}\\
s_{12}s_{23}-c_{12}c_{23}s_{13}\text{e}^{i\delta}&-c_{12}s_{23}-s_{12}c_{23}s_{13}\text{e}^{i\delta}&c_{23}c_{13}
\end{pmatrix}\times
\begin{pmatrix}
1&0&0\\
0&\text{e}^{\frac{i\alpha_{21}}{2}}&0\\
0&0&\text{e}^{\frac{i\alpha_{31}}{2}}
\end{pmatrix}\,,
\end{equation}
%%%%%%%%%%%%%%%%%%
%
where $c_{ij} \equiv \cos\theta_{ij}$, $s_{ij} \equiv \sin\theta_{ij}$,
$\delta$ is the Dirac CP violation (CPV) phase,
while $\alpha_{21}$ and $\alpha_{31}$ are the two Majorana
CPV phases \cite{Bilenky:1980cx}.
In the numerical analysis that follows, we will use
the values of the three neutrino mixing angles
$\theta_{12}$, $\theta_{23}$ and  $\theta_{13}$, and the two neutrino mass squared differences obtained in the global neutrino oscillation data analysis performed in \cite{Esteban_2020} and quoted in Table \ref{tab:PMNSparams}. It follows from \cite{Esteban_2020}, in particular,
that the $3\sigma$ allowed interval of values
of the Dirac CPV phase $\delta$ is rather large.
Furthermore, the Majorana phases $\alpha_{21}$ and $\alpha_{31}$ cannot be constrained by the  neutrino oscillation experiments.
Thus, we will treat the Dirac and Majorana CPV phases as free parameters.

%%%%%%%%%%%%%%%%%%%%%%%%%%%%%%%%%%%%%%%%%%
\newcolumntype{C}{@{}>{\centering\arraybackslash}X}
\begin{table}
\centering
\begin{tabularx}{\textwidth}{@{}CCCCCC@{}}
\hline
\hline
\addlinespace[0.3ex]
    \rowcolor[gray]{.95}
    \multicolumn{6}{c}{\bf Best Fit Values of the Neutrino Parameters}\\
    \hline
    \addlinespace[0.3ex]
    $\theta_{12}$ & $\theta_{13}$ & $\theta_{23}$ & $\delta$ & $\Delta m_{21}^2$ & $\Delta m_{31(32)}^2$\\
    \addlinespace[0.3ex]
    $(^\circ)$  & $(^\circ)$ & $(^\circ)$ & $(^\circ)$ & $(10^{-5}\,\text{eV}^2)$ & $(10^{-3}\,\text{eV}^2)$\\
    \hline
    \addlinespace[0.3ex]
    33.44 & 8.57 & 49.2 & 197 & 7.42 & 2.517\\
    \hline
\end{tabularx}
\caption{The best fit values of
the three neutrino mixing angles $\theta_{12}$, $\theta_{13}$, $\theta_{23}$ and the two neutrino mass squared differences in the case of light neutrino mass spectrum with NO \cite{Esteban_2020}. The best fit value for the Dirac phase $\delta$ is also reported for completeness, even though in our analysis we treat it as a free parameter.}
\label{tab:PMNSparams}
\end{table}

The quantities $(RV)_{\ell j}$ in Eq.~(\ref{eq:nulnuiNj})
determine the strength of the CC and NC weak interaction couplings of the heavy Majorana neutrinos $N_j$ to the $W^\pm$ bosons and
the charged lepton $\ell$, and to the $Z^0$ boson and
the LH flavour neutrino $\nu_{\ell L}$,
$\ell=e,\mu,\tau$ in the weak interaction Lagrangian:
%%%%%%%%%%%%%%%%%%%%%%%%%%%%%%%%%%%%%%%%%%%
\begin{eqnarray}
 \mathcal{L}_\text{CC}^N &=& -\,\frac{g}{2\sqrt{2}}\,
\bar{\ell}\,\gamma_{\alpha}\,(RV)_{\ell j}(1 - \gamma_5)\,N_{j}\,W^{\alpha}\;
+\; {\rm h.c.}\,
\label{eq:NCC},\\
 \mathcal{L}_\text{NC}^N &=& -\frac{g}{4 c_{w}}\,
\overline{\nu_{\ell L}}\,\gamma_{\alpha}\,(RV)_{\ell j}\,(1 - \gamma_5)\,N_{j}\,Z^{\alpha}\;
+\; {\rm h.c.}\,,
\label{eq:NNC}
\end{eqnarray}
%%%%%%%%%%%%%%%%%%%%%%%%%%%%%%%%%%%%%%
%
where $c_w \equiv \cos\theta_w$, $\theta_w$ being the weak mixing angle.

The magnitude of the couplings $(RV)_{\ell j}$
in the region of the parameter space
of successful leptogenesis is crucial for the possibility
to test the low-scale leptogenesis scenarios.

Equation (\ref{eq:seesanuMajM})
allows to relate the matrix of the neutrino Yukawa couplings
$Y$ and the matrix $U$\cite{Casas:2001sr}.
In the diagonal mass basis we are using, this relation has the form (Casas-Ibarra parametrisation):
%%%%%%%%%%%%%%%%%%%%%%%%%%%%%%%%%%%%%
\be
Y =
 i\, \dfrac{\sqrt{2}}{v}\,U\, \sqrt{\hat{m}_\nu}\,O^T\sqrt{\hat{M}}\,,
\label{CI}
\ee
%%%%%%%%%%%%%%%%%%%%%%%%%%%%%%%%%%%%%%
%
where $O$ is a complex orthogonal matrix, $O^T\,O = O\,O^T = I$
and $\hat{M} = {\rm diag}(M_1,M_2,M_3)$.
The usual parametrisation for the matrix $O$, e.g.~adopted in
\cite{Granelli:2020ysj, Klaric:2020lov, Klaric:2021cpi}, is that given
in terms of three Euler complex angles
$\theta_j=\omega_j+i\xi_j$, with $j=1,2,3$ and
$\omega_j,\;\xi_j\in \mathds{R}$ for any $j$, and reads:
%%%%%%%%%%%%%%%%%%%%%%%%%%%%%%%%
\begin{equation}
\label{eq:Euler}
O
=\begin{pmatrix}
c_2c_3 & c_2s_3 & s_2\\
-s_1s_2c_3-c_1s_3 & -s_1s_2s_3+c_1c_3 & s_1c_2\\
-c_1s_2c_3+s_1s_3 & -c_1s_2s_3-s_1c_3 & c_1c_2
\end{pmatrix},
\end{equation}
%%%%%%%%%%%%%%%%%%%%%%%%%%%%%%%
%
where $s_{j} \equiv \sin(\theta_j)$ and $c_{j} \equiv \cos(\theta_j)$.
An equivalent alternative parametrisation was utilised
in \cite{Drewes:2021nqr}.
It has the form:
%%%%%%%%%%%%%%%%%%%%%%%%%%%5
\begin{equation}
\label{eq:alt_O}
O = \left(O_\nu R_C O_N\right)^T,
\end{equation}
%%%%%%%%%%%%%%%%%%%%%%%%%
%
where $O_\nu = O_\nu^{(13)}O_\nu^{(23)}$ and
$O_N = O_N^{(23)}O_N^{(13)}$ represent products of
real rotations in the 1-3 and 2-3 planes,
while $R_C = R_C^{(12)}$ describes a rotation by a complex angle in
the 1-2 plane. This parametrisation proves convenient in the three RH (heavy Majorana)
neutrino case since it involves just one complex angle
(in $R_C$), denoted as $\theta_C$ in what follows.

The $O$-matrix defined above have det$(O) = 1$.
Often, in the literature on the subject, the factor
$\varphi = \pm 1$ is included in the definition of certain elements
of $O$ to allow for the both cases det$(O) = \pm 1$.
We will work with the matrix in  Eq.~\eqref{eq:alt_O},
but extend the range of the Majorana phases $\alpha_{21(31)}$
from  $[0, 2\pi]$ to $[0, 4\pi]$,
which effectively accounts for both cases of $\text{det}(O) = \pm\, 1$
\cite{Molinaro:2008rg}. In this way,
the same full set of $O$ and Yukawa matrices is considered.

From the results obtained in \cite{Drewes:2021nqr} in the three RH neutrino case with quasi-degenerate heavy Majorana neutrinos
for $M_{1,2,3} \cong M \leq 70$ TeV,
it follows, as we have already briefly discussed,
that one can have successful leptogenesis
for either NH or QD light neutrino mass spectrum,
and for $M$ in the ranges $1.7~{\rm GeV} - 70$ TeV
and $50~{\rm MeV} - 70$ TeV in the cases of TIA and VIA, respectively.
In the region of viable leptogenesis, the observable
quantity related to the heavy Majorana neutrino couplings,
$\sum_{\ell j} |(RV)_{\ell j}|^2$,
varies in a wide range, having relatively large values
accessible to low-energy experiments other than, for example, SHiP and those at the discussed FCC-ee collider.
For $m_1 = 0$ (NH spectrum) and
$M = 100$ GeV (70 TeV), for example,
as was reported in \cite{Drewes:2021nqr},
${\rm max}(\sum_{\ell j} |(RV)_{\ell j}|^2) \cong 0.1~(10^{-5})$.
The value of the observable
 $\sum_{\ell, i}|(RV)_{\ell i}|^2$ of interest
exhibits a relatively weak dependence
on the Dirac and Majorana phases, mild dependence on the
Casas-Ibarra real angles of the parametrisation in Eq.~\eqref{eq:alt_O}
and strong dependence on the imaginary part of ${\theta_C}$.
\\

%%%%%%%%%%%%%%%%%%%%%%%%%%%%
\section{Low-Energy Phenomenology:
 Limits and Prospective Tests by cLFV Experiments}
\label{sec:cLFV}
%%%%%%%%%%%%%%%%%%%%%%%%%%%%
%
The low-energy phenomenology of the considered type I seesaw
scenario has been investigated, e.g., in
\cite{Ibarra:2010xw,Ibarra:2011xn,Dinh:2012bp,Penedo:2017knr}.
The CC and NC couplings in Eqs.~\eqref{eq:NCC} and \eqref{eq:NNC}
can induce (via one-loop diagrams with exchange of virtual
$N_{1,2,3}$) charged lepton flavour violating (cLFV) processes
$\mu^\pm \to e^\pm + \gamma$, $\mu^\pm \to e^\pm + e^+ + e^-$,
$\mu - e$ conversion in nuclei, etc.
\cite{Petcov:1976ff,Bilenky:1977du}.

The most stringent upper limits
on the rates of these processes have
been obtained in experiments with muons.
The best experimental limits on $\mu \rightarrow e\gamma$ and
$\mu \rightarrow eee$ decay branching ratios,
$\textrm{BR}(\mu \rightarrow e \gamma)$ and
$\textrm{BR}(\mu \rightarrow eee)$, and on the relative
$\mu -  e$ conversion cross section in a nucleus ${}_{Z}^{A}\textrm{X}$, $\textrm{CR}(\mu\,{}^{A}_{Z}\textrm{X} \rightarrow e\,{}^{A}_{Z}\textrm{X})$ ($Z$ and $A$ are the atomic and mass numbers, respectively),
have been reported by the MEG \cite{MEG:2016leq},
SINDRUM \cite{Bellgardt:1987du}
and SINDRUM II \cite{Dohmen:1993mp,Bertl:2006up} Collaborations:
%%%%%%%%%%%%%%%%%%%%%%%%%
\begin{align}
\textrm{BR}(\mu \rightarrow e\gamma) \,&<\, 4.2 \times 10^{-13}~\textrm{(90\% C.L.)}\,,
\label{eq:MEG} \\
\textrm{BR}(\mu \rightarrow eee) \,&<\, 1.0 \times 10^{-12}~\textrm{(90\% C.L.)}\,,
\label{eq:SINDRUM} \\
\textrm{CR}(\mu\,{}_{22}^{48}\textrm{Ti} \rightarrow e\,{}_{22}^{48}\textrm{Ti}) \,&<\, 4.3 \times 10^{-12}~\textrm{(90\% C.L.)}\,,
\label{eq:SINDRUMII} \\
\textrm{CR}(\mu\,{}_{~79}^{197}\textrm{Au} \rightarrow e\,{}_{~79}^{197}\textrm{Au}) \,&<\, 7.0 \times 10^{-13}~\textrm{(90\% C.L.)}\,.
\label{eq:SINDRUMII2}
\end{align}
%%%%%%%%%%%%%%%%%%%%%
%
The planned MEG II update of the MEG experiment~\cite{MEGII:2018kmf}
aims at reaching sensitivity to
$\textrm{BR}(\mu \rightarrow e\gamma) \simeq 6 \times 10^{-14}$.
The sensitivity to $\textrm{BR}(\mu \rightarrow eee)$ is
planned to be increased by up to three (four) orders of magnitude
to  $\textrm{BR}(\mu \rightarrow eee) \,\sim\, 10^{-15}~(10^{-16})$
with the realisation of Phase I (Phase II)
of the  Mu3e Project~\cite{Arndt:2009}.
The Mu2e~\cite{Bartoszek:2015} and COMET~\cite{Abramishvili:2020}
collaborations studying $\mu - e$ conversion in aluminium plan to
reach sensitivity to
$\textrm{CR}(\mu\,{}_{13}^{27}\textrm{Al} \rightarrow e\,{}_{13}^{27}\textrm{Al}) \sim\,
6 \times 10^{-17}$.
The planned PRISM/PRIME
experiment~\cite{Barlow:2011zza} aims at
a dramatic increase of sensitivity to
the $\mu - e$ conversion rate in titanium,
allowing to probe values as small as
$\textrm{CR}(\mu\,{}_{22}^{48}\textrm{Ti} \rightarrow e\,{}_{22}^{48}\textrm{Ti}) \,\sim\, 10^{-18}$,
an improvement  by six orders of magnitude of the current bound
given in Eq.~\eqref{eq:SINDRUMII}.

The predictions of the seesaw model under discussion, e.g., for the
rates of the $\mu \rightarrow e\gamma$ and $\mu \rightarrow eee$ decays and
$\mu -  e$ conversion in nuclei, as can be shown, depend on the quantity
$|\sum_{i=1,2,3}(RV)_{\mu i}^* (RV)_{e i}|^2$,
and, for $|M_i - M_j| \ll M_k$, $i\neq j = 1,2,3$, $k=1,2,3$,
on the mass $M_{1,2,3} \simeq M$ of
the heavy Majorana neutrinos $N_{1,2,3}$.
The expressions for $\textrm{BR}(\mu \rightarrow e \gamma)$,
$\textrm{BR}(\mu \rightarrow eee)$ and $\textrm{CR}(\mu\,{}^{A}_{Z} \textrm{X} \rightarrow e\,{}^{A}_{Z}\textrm{X})$ in the case of interest can be easily
obtained from those given in
Refs.~\cite{Ibarra:2011xn,Dinh:2012bp,Alonso:2012ji} and we are not going
to reproduce them here.
Let us add that the rates of the cLFV decays of
the $\tau$ lepton are proportional to the product of couplings
$|\sum_{j=1,2,3} (RV)^*_{\tau j}(RV)_{\ell' j}|^2$, $\ell' = e,\mu$.
However, the current constraints and the prospective
improvements of the sensitivity of the experiments on
cLFV decays of $\tau^\pm$  are
respectively less stringent and not so significant
as in the case of experiments on cLFV processes with $\mu^\pm$
and we are not going to consider them here.

\begin{figure}
    \centering
\includegraphics[width=0.76\textwidth]{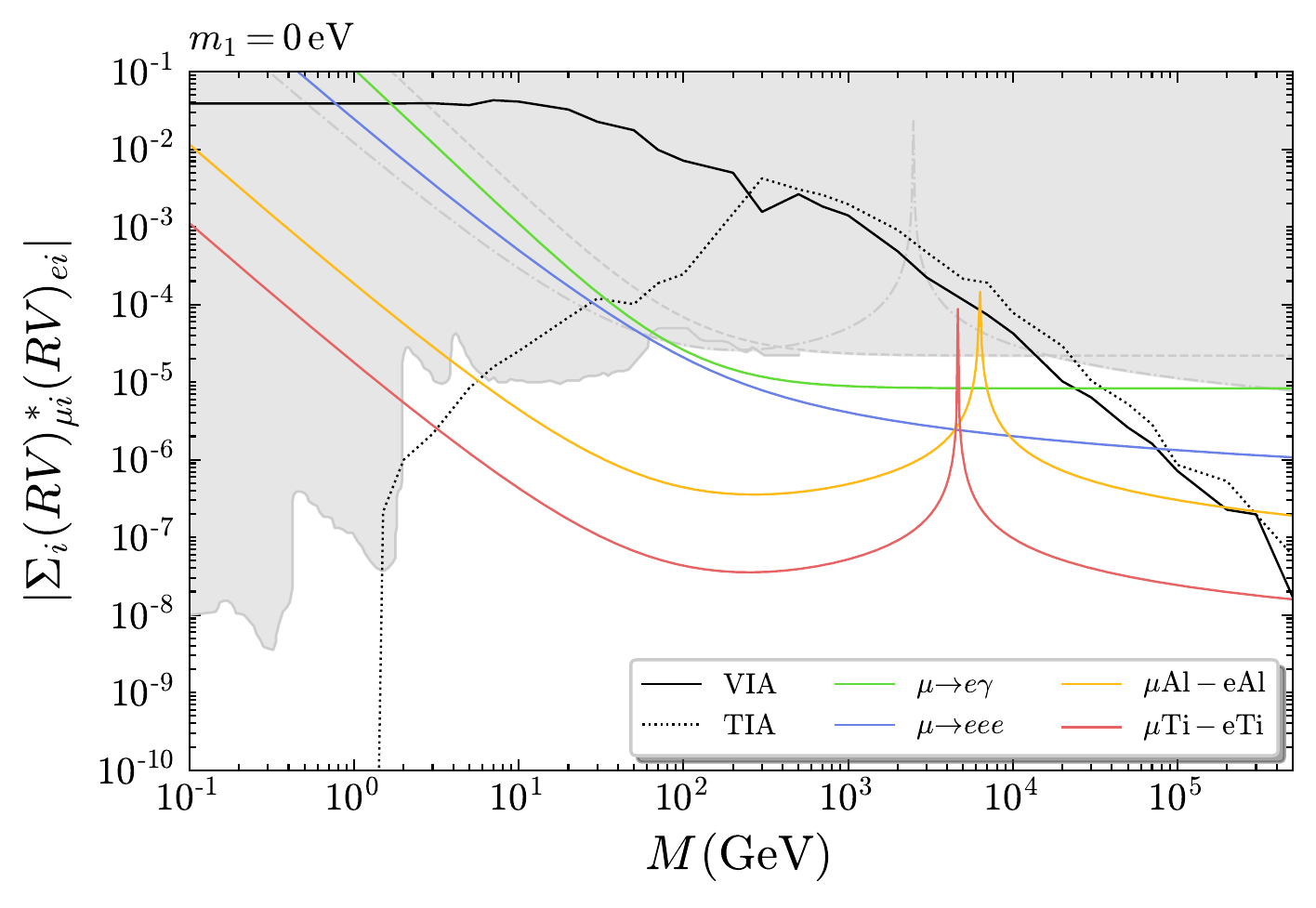}
\includegraphics[width=0.76\textwidth]{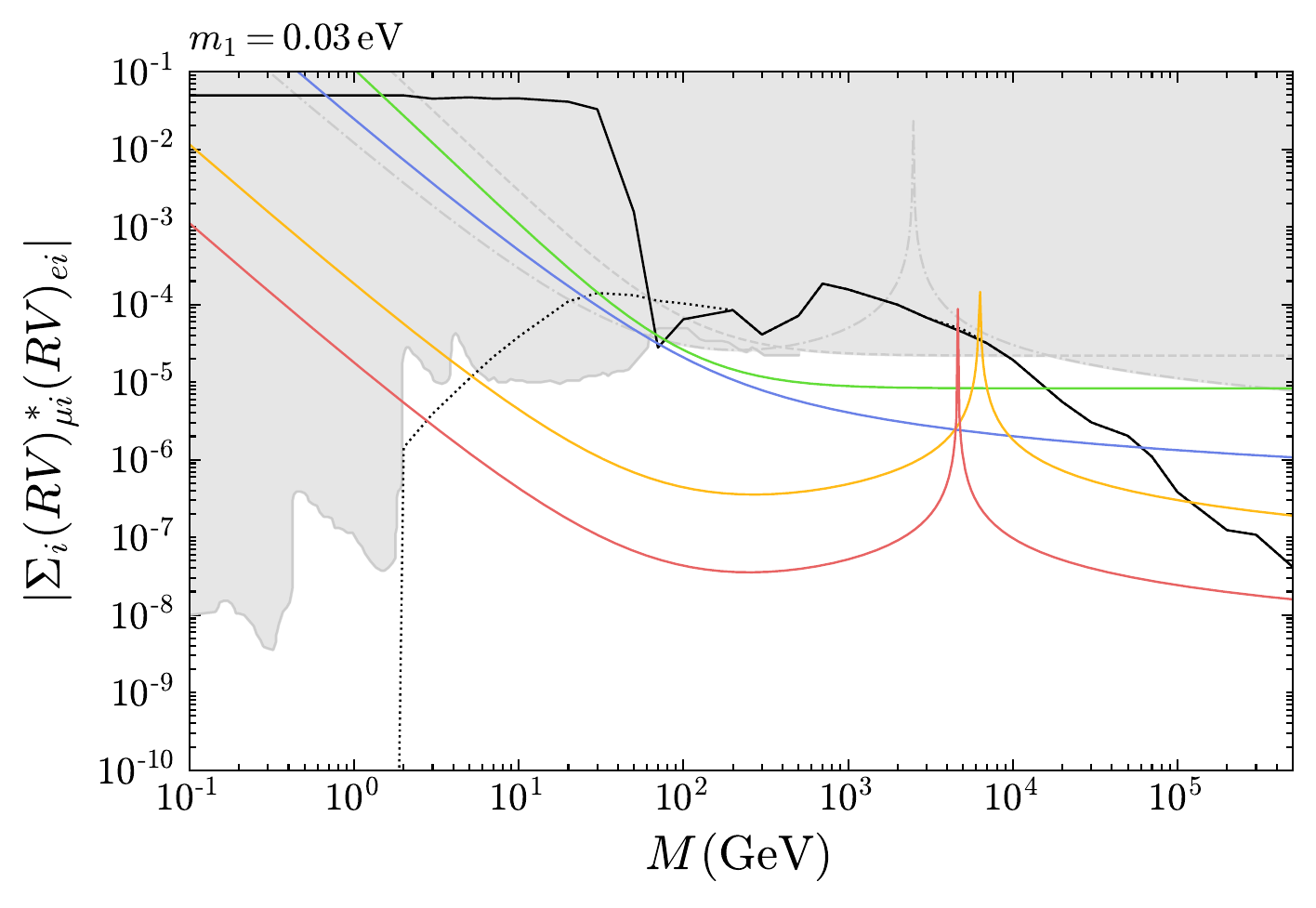}
    \caption{The region in the
$|\sum_{i=1,2,3} (RV)^*_{\mu i} (RV)_{e i}| - M$ plane
of successful low-scale leptogenesis in the case
of NH light neutrino mass spectrum with $m_1 = 0$ (top panel) and for NO spectrum with $m_1 = 0.03$ eV (bottom panel).
The solid and dotted black curves are the constraints from successful
leptogenesis in the VIA and TIA cases, respectively. The grey region with
solid contour that extends to $M\sim 500$ GeV is excluded by low-energy
experiments as shown in \cite{Chrzaszcz_2020}, that with dashed and dot-dashed
contours are excluded by the current upper limits
$\text{BR}(\mu \rightarrow e\gamma) < 4.2 \times 10^{-13}$ \cite{MEG:2016leq}
and $\text{CR}(\mu\,{}_{13}^{27}\text{Au} \rightarrow e\,{}_{13}^{27}\text{Au})
< 7 \times 10^{-13}$ \cite{Bertl:2006up}, respectively.
The green, blue, yellow and red lines correspond, from top to bottom, to the
sensitivities of the upcoming experiments on $\mu^\pm \to e^\pm + \gamma$, $\mu^\pm \to e^\pm + e^+ + e^-$ decays and on $\mu - e$ conversion in aluminium and titanium. See the text for further details.}
\label{fig:cLFV}
\end{figure}

In the region of viable leptogenesis, the quantity of interest
 $|\sum_{i=1,2,3}(RV)^*_{\mu i}(RV)_{e i}|$
can be as large as $10^{-1}$ (see  Fig.~\ref{fig:cLFV}), which opens up
the possibility to test the low-scale leptogenesis scenario with
three quasi-degenerate heavy Majorana neutrinos in
experiments on cLFV with $\mu^\pm$.
Indeed, consider as an example the experiments
on $\mu \rightarrow e \gamma$ decay.
The $\mu \rightarrow e \gamma$ decay branching ratio
is given by \cite {Ibarra:2011xn} (see also
\cite{Petcov:1976ff,Bilenky:1977du,Cheng:1980tp}):
%%%%%%%%%%%%%%%%%%%%%%%%%%%%%%%%%%
\begin{eqnarray}
\text{BR}(\mu\to e\gamma) =
\frac{\Gamma(\mu\to e+\gamma)}{\Gamma(\mu\to e+\nu_{\mu}+\overline{\nu}_{e})}
&=&
\frac{3\alpha_{\rm em}}{32\pi}\,|T|^{2}\,,
\label{eq:Bmutoeg1}
\end{eqnarray}
%%%%%%%%%%%%%%%%%%%%%%
%
where $\alpha_{\rm em}$ is the fine structure constant and
%%%%%%%%%%%%%%%%%%%%%%%%%%%%
\begin{equation}
	T\;\cong\; \,
\left[ G(X) - G(0)\right]\sum_{i=1,2,3}(RV)_{\mu i}^{*}\, (RV)_{ei}\,.
\label{eq:T2}
\end{equation}
%%%%%%%%%%%%%%%%%%%%%%%%%%%
%
Here, $G(X)$ is a loop integration function,
$X \equiv(M/M_{W})^{2}$ and we have
taken into account that the differences between
$M_1$, $M_2$ and $M_3$ are negligibly small, with $M_{1,2.3}\cong M$. The function $G(X)$ is monotonic
\footnote{The explicit analytic expression for the function $G(X)$
 can be found in \cite{Ibarra:2011xn}.
}
and takes values in the interval
$[4/3,10/3]$, with
$G(X)\cong10/3-X$ for $X\ll 1$. At, e.g., $M = M_{W}$ ($M = 1000$ GeV)
we have
$G(X) - G(0) = -\,0.5\, (\simeq -1.9)$. It is not difficult to show,
using these values of $G(X) - G(0)$ and
Eqs.~\eqref{eq:Bmutoeg1} and \eqref{eq:T2},
that the MEG II experiment aiming to probe $\text{BR}(\mu \rightarrow e\gamma)$ down to $6 \times 10^{-14}$,
will be sensitive for $M = M_{W}$ ($M = 1000$ GeV) to values of
$|\sum_{i=1,2,3}(RV)_{\mu i}^{*}\, (RV)_{ei}| \gtrsim 3.3\times 10^{-5}~(8.9 \times 10^{-6})$.
This is approximately by 1 to 3  orders of magnitude
smaller than the maximal value of
$|\sum_{i=1,2,3}(RV)_{\mu i}^{*}\, (RV)_{ei}|$ at $M = M_{W}$ ($M = 1000$ GeV)
for which we can have successful
low-scale leptogenesis in the scenario
with three quasi-degenerate in mass heavy Majorana neutrinos
in the TIA and VIA cases.

Even smaller values of $|\sum_{i=1,2,3}(RV)_{\mu i}^{*}\, (RV)_{ei}|$
can be probed in the Mu3e experiment~\cite{Arndt:2009},
planning to reach sensitivity to
$\textrm{BR}(\mu \rightarrow eee) \,\sim\, 10^{-15}~(10^{-16})$
and especially in the upcoming
Mu2e~\cite{Bartoszek:2015}, and COMET~\cite{Abramishvili:2020}
experiments on  $\mu - e$ conversion in aluminium, aiming
ultimately to be sensitive to $\textrm{CR}(\mu\,{}_{13}^{27}\textrm{Al} \rightarrow e\,{}_{13}^{27}\textrm{Al}) \sim\,
6 \times 10^{-17}$.
Values as small as  $|\sum_{i=1,2,3}(RV)_{\mu i}^{*}\, (RV)_{ei}|\sim 10^{-7}$
at $M\sim 100$ GeV can be probed in planned PRISM/PRIME
experiment~\cite{Barlow:2011zza}, aiming at
an impressive increase of sensitivity to
the $\mu - e$ conversion rate in titanium to
$\textrm{CR}(\mu\,{}_{22}^{48}\textrm{Ti} \rightarrow e\,{}_{22}^{48}\textrm{Ti}) \,\sim\, 10^{-18}$.

In order to obtain the region of viable leptogenesis in terms of the
cLFV observable quantities, we solve the density matrix equations
from~\cite{Klaric:2020lov,Drewes:2021nqr}, and scan the parameter space
for the largest allowed values of $|\sum_{i=1,2,3}(RV)_{\mu i}^{*}\, (RV)_{ei}|$.
In Fig.~\ref{fig:cLFV} we show the regions of viable low-scale leptogenesis
in the considered scenario in the
$|\sum_{i=1,2,3}(RV)_{\mu i}^{*}\, (RV)_{ei}| - M$
plane for  $|\sum_{i=1,2,3}(RV)_{\mu i}^{*}\, (RV)_{ei}| \geq 10^{-11}$ and $M$
in the interval $M = (0.1 - 5\times 10^{5})$ GeV
% $M = (0.1 - 5\times 10^{5})$ GeV
in the TIA and VIA cases (regions below the dotted and solid black lines,
respectively).  The light neutrino mass spectrum is assumed to be with NO.
The lightest neutrino mass is set to $m_1 = 0$ (top panel) and
$m_1 = 0.03$ eV (bottom panel). The subregion which is excluded by the
current low-energy data \cite{Chrzaszcz_2020}, including the current upper
limitations on
$\textrm{BR}(\mu \rightarrow e \gamma)$ and on
$\textrm{CR}(\mu\,{}_{~79}^{197}\textrm{Au} \rightarrow e\,{}_{~79}^{197}\textrm{Au})$ given in
Eqs.~\eqref{eq:MEG} and \eqref{eq:SINDRUMII2}, is shown in grey.
The green, blue, yellow and red lines represent, from top to bottom,
the prospective sensitivities of the planned experiments
on $\mu \rightarrow e\gamma$ and $\mu \rightarrow eee$
decays, as well as on $\mu - e$ conversion in aluminium and titanium
\footnote{The spikes in the curves related to $\mu - e$ conversions,
appearing for different RH neutrino masses in relation to the considered
nucleus, are present because the relative rates of the processes,
calculated at leading (one-loop) order and neglecting the differences
between the masses of $N_{1,2,3}$,
go through zero \cite{Dinh:2012bp,Alonso:2012ji,Ilakovac:2009jf}.
}.
As the two figures clearly indicate,
the planned experiments on cLFV with $\mu^\pm$
(i.e., on $\mu$LFV)
can probe directly significant region of the
leptogenesis parameter space, which cannot be explored
by any other experiments.
More specifically, the future MEG II and Mu3e
experiments on $\mu \rightarrow e \gamma$
and $\mu \rightarrow eee$ decays will probe the currently
allowed leptogenesis regions, which extend respectively
from $M\cong 90$ GeV to $M\cong 2\times 10^4$ GeV
and from $M\cong 60$ GeV to $M\cong 7\times 10^4$ GeV in the VIA case and
to slightly larger values in the TIA case;
they will probe values of
the parameter $|\sum_{i=1,2,3}(RV)_{\mu i}^{*}\, (RV)_{ei}|$
down to $8\times 10^{-6}$ and $1.5\times 10^{-6}$.
Except for a narrow region in the vicinity of the
spike at 6.0 TeV, in the VIA (TIA) case
% and 4.5 TeV,
the upcoming experiments on $\mu - e$ conversion in aluminium
Mu2e~\cite{Bartoszek:2015} and COMET~\cite{Abramishvili:2020}
will probe the allowed leptogenesis
region within the interval $M\cong (4~(6) - 3\times 10^5$ GeV
and values of $|\sum_{i=1,2,3}(RV)_{\mu i}^{*}\, (RV)_{ei}|$
down to $2\times 10^{-7}$,
while the planned experiment with higher sensitivity
on  $\mu - e$ conversion in titanium  PRISM/PRIME~\cite{Barlow:2011zza}
will test (apart from a narrow interval around the spike at
4.5 TeV) the leptogenesis region
in the range of $M\cong 2~(3) - 5\times 10^5$ GeV and values of
$|\sum_{i=1,2,3}(RV)_{\mu i}^{*}\, (RV)_{ei}|$ as small as $1.6\times 10^{-8}$.
%%%%%%%%%%%%%%%%%%%%%%%%%%%%%%%%%%%%%
%%%%%%%%%%%%%%%%%%%%%%%%%%%%%%%%%%%%%
%
If any of the considered $\mu$LFV experiments finds a positive result,
that will serve also as an indication in favour of the considered
low-scale leptogenesis scenario with three (RH)
quasi-degenerate in mass heavy Majorana neutrinos.
From the data on the rate of the observed process one would
determine the values of $M$ and $|\sum_{i=1,2,3}(RV)_{\mu i}^{*}\, (RV)_{ei}|$
(with certain uncertainties).
That will allow to make specific predictions for the
rates for the other two processes, which, if confirmed experimentally,
would constitute further evidence for the discussed low-scale leptogenesis scenario with three RH neutrinos based on the type I seesaw mechanism of neutrino mass generation.

\begin{figure}
    \centering
    \includegraphics[width=0.76\textwidth]{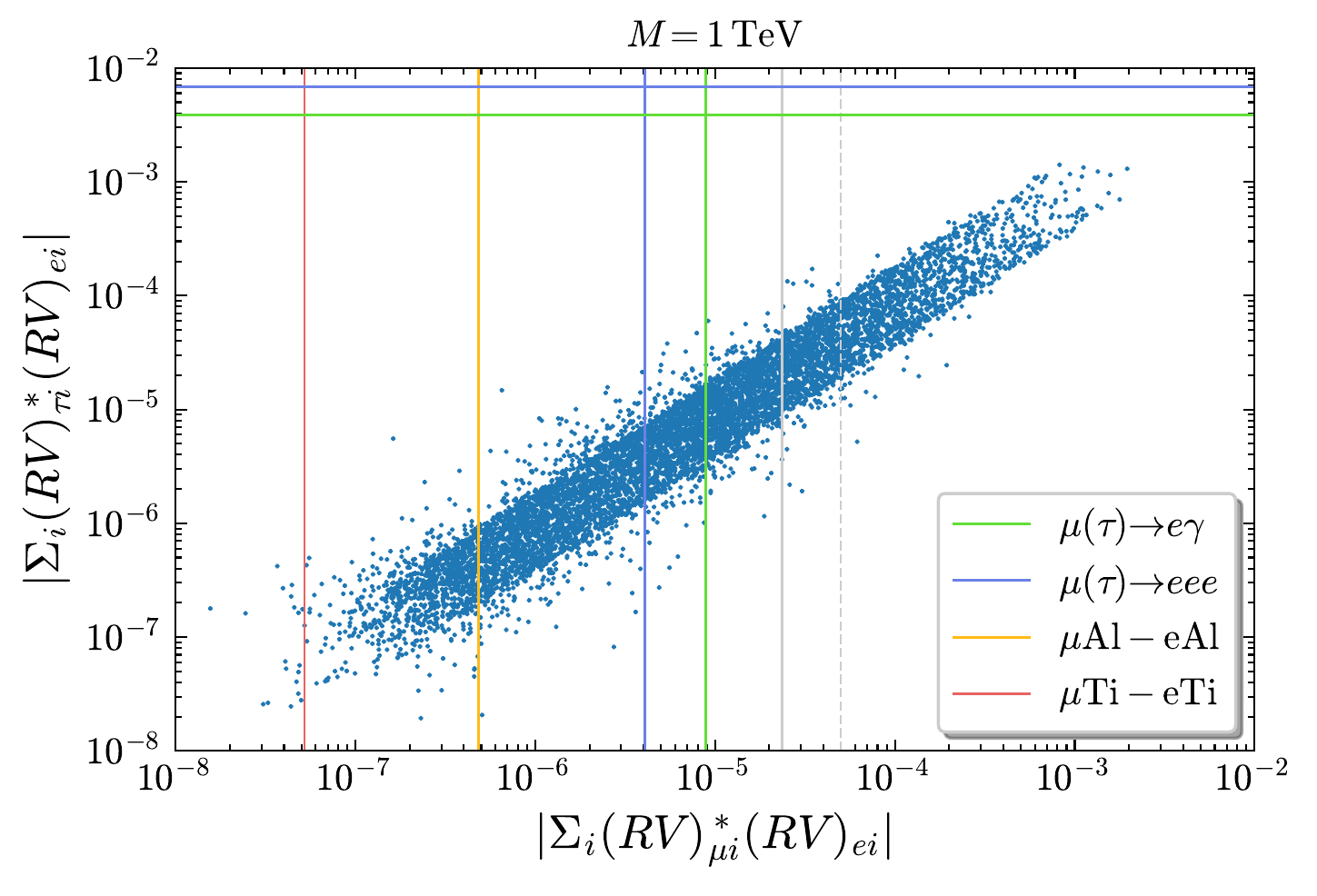}
    \includegraphics[width=0.76\textwidth]{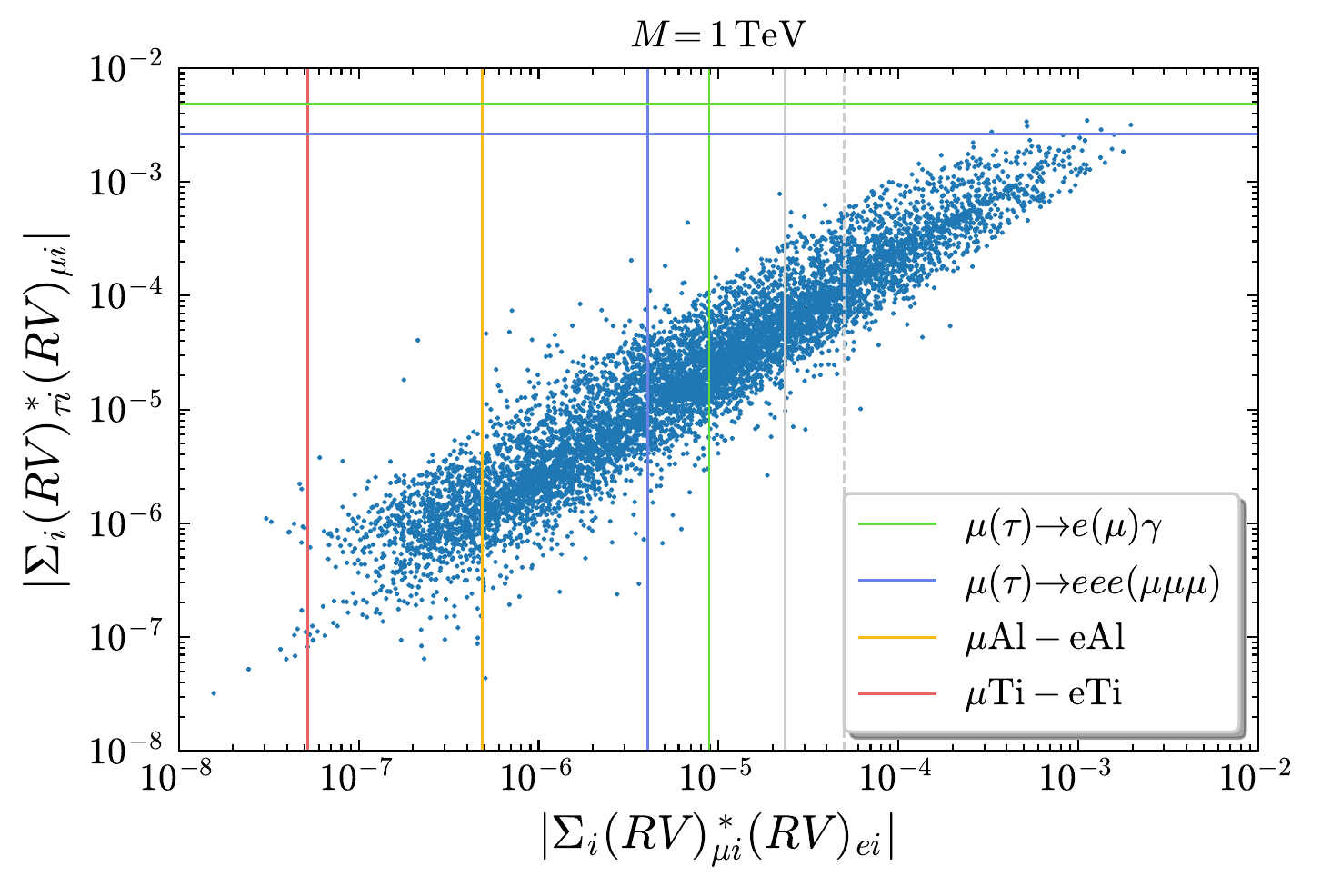}
    \caption{ We show in the top (bottom) panel of the figure the points 
in the $|\sum_i (RV)_{\mu i}^* (RV)_{e i}| - |\sum_i (RV)_{\tau i}^* (RV)_{e(\mu) i}|$ 
plane for which we find viable leptogenesis for $M=1\,\text{TeV}$ and $m_1 = 0$ (NH light neutrino mass spectrum). 
The vertical grey lines are the upper limits
on $|\sum_i (RV)_{\mu i}^* (RV)_{e i}|$ implied by the current limits 
$\text{BR}(\mu \rightarrow e\gamma) < 4.2 \times 10^{-13}$ (solid) and 
$\text{CR}(\mu\text{Au} \rightarrow e\text{Au}) < 7 \times 10^{-13}$ (dashed). 
The green, blue, yellow and red vertical lines, from right to left, 
correspond to the sensitivities on $|\sum_i (RV)_{\mu i}^* (RV)_{e i}|$ of the 
upcoming $\mu$LFV experiments on $\mu \to e\gamma$, $\mu \to eee$, $\mu - e$ 
in aluminium and $\mu - e$ in titanium, planing to reach, respectively, 
$\text{BR}(\mu \rightarrow e\gamma) \sim 6 \times 10^{-14}$, 
$\text{BR}(\mu \rightarrow eee) \sim 10^{-15}$, 
$\text{BR}(\mu \text{Al}\rightarrow e\text{Al}) \sim 6 \times 10^{-17}$ and 
$\text{BR}(\mu \text{Ti} \rightarrow e\text{Ti}) \sim 10^{-18}$. 
The horizontal blue and green lines in the top (bottom) panels are, 
from top (bottom) to bottom (top), 
the sensitivities on $|\sum_i (RV)_{\tau i}^* (RV)_{e(\mu) i}|$ of upcoming 
experiments on $\tau \to eee(\mu\mu\mu)$ and $\tau \to e(\mu)\gamma$, 
planning to reach sensitivity to
$\text{BR}(\tau \rightarrow eee(\mu\mu\mu)) \sim 5 \times 10^{-10} (7\times 10^{-11})$ and $\text{BR}(\tau \rightarrow e(\mu)\gamma) \sim 2 (3)\times 10^{-9}$ \cite{BELLEIIbook, BELLEII, Workman:2022ynf}.
}
\label{fig:UmuUevsUtauUe}
\end{figure}

 We note that in the region of parameter space of successful leptogenesis, 
the heavy Majorana neutrinos can have sizeable CC couplings 
not only to  the electron and muon, but 
to  the electron, muon and tauon simultaneously.
This is illustrated 
in Fig.~\ref{fig:UmuUevsUtauUe} in which we show
a generic example of points in the leptogenesis parameter space for 
$M=1\,\text{TeV}$ where 
both $\mu$-LFV and $\tau$-LFV processes 
are possible simultaneously and can proceed 
with rates that can be probed in future planned experiments.
%%%%%%%%%%%%%%%%%%%%%%%%%%%%%%%%%%%%%%%%%%%
%%%%%%%%%%%%%%%%%%%%%%%%%%%%%%%%%%% 
%

%%%%%%%%%%%%%%%%%%%%%%%%%%%%%%
%
\section{Summary}
%
%%%%%%%%%%%%%%%%%%%%%%%%%%%%%%%%%%5
%

To summarise, we have shown that the upcoming and planned experiments
on charged lepton flavour violation with $\mu^\pm$, MEG II on
the $\mu \rightarrow e\gamma$ decay,
Mu3e on $\mu \rightarrow eee$ decay,
Mu2e and COMET on $\mu - e$ conversion in aluminium
and PRISM/PRIME on $\mu - e$ conversion in titanium,
can probe directly significant regions of the
viable parameter space of  low-scale leptogenesis based on the
type I seesaw mechanism with three quasi-degenerate in mass heavy
Majorana neutrinos $N_{1,2,3}$, and thus test this attractive
leptogenesis scenario with a potential for a discovery.
The BELLE II experiments on 
$\tau \to eee(\mu\mu\mu)$ and $\tau \to e(\mu)\gamma$
also can probe a part of the leptogenesis parameter space, 
although a relatively small one.
We are looking forward to the results of
these very important experiments on beyond the Standard Model physics.

\section*{Acknowledgements}
We thank Patrick D.~Bolton for useful discussions on aspects of low-energy tests of the low-scale leptogenesis scenarios discussed in the present article.
We also thank Kevin Alberto Urquía Calderón, Oleg Ruchayskiy and Inar Timiryasov for informing us about their upcoming related work.
The work of A.G.~and S.T.P.~was supported in part by the European Union's Horizon 2020 research and innovation programme under the Marie Skłodowska-Curie grant agreement No.~860881-HIDDeN, and by the Italian INFN program on Theoretical Astroparticle Physics. S.T.P.~acknowledges partial support from the World Premier International Research Center Initiative (WPI Initiative, MEXT), Japan.
J.K.~acknowledges the support of the Fonds de la Recherche Scientifique - FNRS under Grant No.~4.4512.10.
Computational resources have been provided by the Consortium des Équipements de Calcul Intensif (CÉCI), funded by the Fonds de la Recherche Scientifique de Belgique (F.R.S.-FNRS) under Grant No.~2.5020.11 and by the Walloon Region.
J.K. and S.T.P. acknowledge the support of the Mainz Institute of Theoretical Physics (MITP) and the University of Naples ``Federico II''  during the final stages of this work at the Program on \emph{Neutrinos, flavour and beyond}, Capri, Italy, June 6-18 2022.
MITP is supported by the Cluster of Excellence {\em Precision Physics, Fundamental Interactions, and Structure of Matter} (PRISMA$^+$ EXC 2118/1) funded by the German Research Foundation (DFG) within the German Excellence Strategy (Project ID 39083149).

%\bibliography{TestingLG}{}
%\bibliographystyle{JHEP}

\providecommand{\href}[2]{#2}\begingroup\raggedright\endgroup

\end{document}